\documentclass[article,twocolumn,showpacs,amsmath,amssymb,floatfix,citeautoscript]{revtex4-1}
\usepackage{graphicx}
\usepackage{dcolumn}
\usepackage{bm}
\usepackage{amsmath}
\usepackage{amssymb}
\usepackage{amssymb, verbatim}
\usepackage[normal]{subfigure}

\usepackage{hyperref}
\hypersetup{
        colorlinks=true,
}

\makeatletter

\begin{document}

\title{Classification of Floquet Statistical Distribution for Time-Periodic Open Systems}

\author{Dong E. Liu}
\affiliation{
Microsoft Research, Station Q, Santa Barbara, CA, 93106, USA
}

\date{\today}

\begin{abstract} 
How to understand the order of Floquet stationary states 
in the presence of external bath coupling and their statistical mechanics is challenging;
the answers are important for preparations and control of those Floquet states.
Here, we propose a scheme to classify the statistical distribution of Floquet states for
time-periodic systems which couple to an external heat bath. If an effective Hamiltonian
and a system-bath coupling operator, which are all time-independent, can be simultaneously 
obtained via a time-periodic unitary transformation, the statistical mechanics of the 
Floquet states is equivalent to the equilibrium statistical mechanics of the effective 
Hamiltonian. In the large driving frequency cases, we also show that the 
conditions of this theorem can be weakened to: the time-period part 
in the system Hamiltonian commutes with the system-bath coupling operator.
A Floquet-Markov approach is applied to numerically compute the Floquet
state occupation distribution of a bosonic chain, and the results agree with 
the theoretical predictions.
\end{abstract}


\maketitle

{\em Introduction}.
It is proposed that time-periodic external fields may be used to engineer exotic quantum phenomena, e.g. quantum phase transitions \cite{eckardt05}, 
topological band insulators \cite{inoue10,lindner11,kitagawaGF11,IadecolaPRL13,Iadecola14,Usaj14,FoaTorres14}, quantum Hall states \cite{Dahlhaus11,Zhou14}, 
Majorana fermions and topological superconductors \cite{jiang11,reynoso12,Liu13,Kundu13,Foster14,PeiWang14,Zheng14}, and so on. 
The scheme of topological classifications are discussed for isolated non-interacting time-periodic systems \cite{kitagawa10,RudnerPRX13}.
Some related experimental signatures are reported recently in a gyromagnetic photonic crystal~\cite{RechtsmanNature13}
and a topological insulator with an intense ultrashort mid-infrared pulse\cite{WangScience13},
(the data in \cite{WangScience13} can also be explained in a different way \cite{Fregoso13,Fregoso14}).
The knowledge of those phenomena mostly comes from a theorem 
applied to time-periodic isolated quantum non-interacting systems (i.e. Floquet theorem), 
and the states in those systems belong to a certain type of
non-equilibrium stationary states--Floquet states \cite{shirley65,sambe73}.
In reality, interactions or heat bath couplings always result in relaxations and
re-distributions in Floquet states.
In that case, one have to understand the statistical mechanics of those Floquet systems.

Thermodynamics and statistical mechanics of a time-periodic quantum system has long been an
elusive topic. Recent studies for time-periodic isolated interacting quantum systems show that
periodic ergodic system will relax to an infinite-temperature distribution in the thermodynamics limit
\cite{DAlessio&Rigol14,Lazarides14,Ponte14}.
For time-periodic open quantum systems (i.e. with an external bath),
some studies based on a Floquet-Markov approach show that the occupation distributions of Floquet states 
have non-trivial behaviors \cite{Kohler97,Hone97,Breuer00,Kohn01,Hone09,Ketzmerick10}:
The occupation distribution has Boltzmann-like weight in some regimes, and
has almost equal probabilities in some other regimes.
Those behaviors are also related to the regular and chaotic behaviors in
Poincare section of the classical phase space \cite{Ketzmerick10}.
Very interestingly, the conserved Noether charge associated with certain continuous
symmetry of the full time-dependent Lagrangian can be used to understand Floquet statistical mechanics \cite{Iadecola13}.
It is clear that in general the statistical
properties of Floquet states do not obey the standard equilibrium statistical mechanics.
When and how do the statistical properties start to deviate
from the standard equilibrium statistical mechanics is still unclear for time-periodic open systems.
We also want to ask whether those non-trivial behaviors depend on
the form of the time-periodic modulation and the system-bath coupling, and when do the open Floquet systems
resemble some equilibrium systems?
Understanding those questions would be helpful for experimental preparations and
control of the Floquet states.

{\em Summary of the main results}.
We study the statistical mechanics of a generic open Floquet system, 
which is modeled by a time-periodic system (the time-independent part has a finite energy band) coupled to a heat bath. 
We find a general classification theorem to classify the statistical properties of Floquet states: 
If both the time-periodic system Hamiltonian and the system-bath coupling operator
can be simultaneously transformed to time-independent forms via 
a time-periodic unitary transformation, the statistical mechanics of the Floquet states,
e.g. the concept of temperature and Floquet occupation statistical distribution,
is the exactly the same as the standard equilibrium statistical mechanics of 
the corresponding time-independent effective Hamiltonian. 
The order of the Floquet states (quasi-energies) can only be understood from
the order of the corresponding effective Hamiltonian eigenvalues.
For large driving frequency $\omega> D$ (where $D$ is the largest energy
scale, e.g. band width), the effective Hamiltonian
can be obtained perturbatively up to certain order of $D/\omega$ \cite{Verdeny13}.
Up to the leading order, the conditions of the theorem---when 
the equilibrium statistical mechanics works---can be weakened to:
$[H_{\rm D},A_{\rm S}]=0$, where
$H_{\rm D}$ is the time-periodic part in the system Hamiltonian and $A_{\rm S}$ is the 
system-bath coupling operator .
On the other hand, if those conditions are not satisfied, Floquet occupation distribution 
does not follow Boltzmann distribution, and one cannot define temperature. 
To test the theorem, we consider a one-dimensional
tight-binding chain with bosons in the presence of a heat bath, and
numerically compute the occupation distribution of the Floquet states.

{\em Theoretical Model: Open Floquet systems}.
We consider a time-periodic quantum system with 
an external heat bath, and the whole system can be modeled in a standard way 
\cite{Feynman&Vernon,caldeira81,LeggettRMP87}
\begin{equation}
 H(t)=H_{\rm S}(t)+H_{\rm B} + H_{\rm SB},
  \label{eq:SB_Hamiltonian}
\end{equation}
where the quantum system is periodic in time $H_{\rm S}(t)=H_{\rm S}(t+T)$ with time period $T$.
The heat bath is modeled by an ensemble of harmonic oscillators  
$H_{\rm B}=\sum_{n}\big( p_n^2/2 m_n +m_n \omega_n^2 x_n^2/2 \big)$. One usually assumes the coupling
between the system and bath is bilinear
\begin{equation}
 H_{\rm SB} = \gamma A_{\rm S} \sum_n c_n x_n,
\end{equation}
where $\gamma$ is the coupling strength and $A_{\rm S}$ is a system-bath coupling operator.
We consider a stationary bath and a time-independent system bath coupling throughout the paper.

Without the heat bath, the solution of the Schrodinger equation for a time-periodic 
Hamiltonian can be obtained from a Floquet theorem \cite{shirley65,sambe73}: The wavefunction
can be factorized 
$|\psi_{\alpha}(t)\rangle=e^{-i\epsilon_{\alpha}t}|\phi_{\alpha}(t)\rangle$ (hereafter $\hbar=1$), where
$\epsilon_{\alpha}$ is called quasi-energy and $|\phi_{\alpha}(t)\rangle=|\phi_{\alpha}(t+T)\rangle$
is called Floquet state. Therefore, one reaches a time-dependent eigenvalue problem 
$[H_{\rm S}(t)-i\partial_t]|\phi_{\alpha}(t)\rangle=\epsilon_{\alpha}|\phi_{\alpha}(t)\rangle$.
With the heat bath, the system-bath coupling induces the transitions and thus relaxation between 
different Floquet states, so we have to consider their statistical properties, e.g. occupation distribution.
 The occupation distribution of some Floquet model systems was studied by using
Floquet-Markov approach \cite{Kohler97,Breuer00,Kohn01,Hone09,Ketzmerick10};
and those studies show that the concepts for equilibrium statistical mechanics
are not generally applicable. 
Therefore, one may ask: Can we find a way to classify the statistical 
distribution via the time-periodic Hamiltonian and the system-bath
coupling operator?

{\em A general classification theorem for Floquet statistical mechanics}.
Since both system Hamiltonian $H_{\rm S}$ and the Floquet states $|\phi_{\alpha}(t)\rangle$
are periodic in time, one can apply the Fourier expansion and rewrite the Floquet
operator $H_{\rm S,F}=H_{\rm S}(t)-i\partial_t$ in an extended matrix form
\begin{equation}
 H_{\rm S,F} = 
 \begin{pmatrix}
  \ddots & \vdots & \vdots & \vdots & \cdots \\
  \cdots & H_{\rm S,0}+\omega & H_{\rm S,1} & H_{\rm S,2} & \cdots \\
  \cdots & H_{\rm S,-1} & H_{\rm S,0} & H_{\rm S,1} & \cdots \\
  \cdots & H_{\rm S,-2} & H_{\rm S,-1} & H_{\rm S,0}-\omega & \cdots \\
  \cdots & \vdots & \vdots & \vdots & \ddots \\
 \end{pmatrix}
 \label{eq:HSF_matrix}
\end{equation}
where the block matrix elements are Fourier series coefficients 
$H_{\rm S}(t)=\sum_n H_{\rm S,n} e^{-i n \omega t}$ and $\omega=2\pi/T$.
If we can find a unitary transition $U_F$ to block-diagonalize $H_{\rm S,F}$ 
\begin{equation}
U_F^{\dagger} H_{\rm S,F} U_F = \rm{Diag} [\cdots,H_{\rm S,eff}+\omega,H_{\rm S,eff},H_{\rm S,eff}-\omega, \cdots],
\label{eq:HeffMF}
\end{equation}
the quasi-energy $\epsilon_{\alpha}$ are related to the eigenvalues $E_{\rm eff,\alpha}$ of
$H_{\rm S,eff}$ via the relation $\epsilon_{\alpha}=\mod (E_{\rm eff,\alpha},\omega)$.
Note that the unitary transition $U_F$ of the matrix Hamiltonian Eq. (\ref{eq:HSF_matrix})
is equivalent to a time-periodic unitary transition $\hat{U}_F(t)$ of $H(t)$ with
$H_{\rm S,eff}=\hat{U}_F(t)^{\dagger}H_{\rm S}(t)\hat{U}_F(t)-i\hat{U}_F(t)^{\dagger}\partial_t \hat{U}_F(t)$. 

Just like $H_{\rm S,F}$, the system-bath coupling can also be written into an extended matrix form
$H_{\rm SB}=\gamma\,\rm{Diag}[\cdots,A_{\rm S},A_{\rm S},A_{\rm S},\cdots]\otimes \sum_n c_n x_n$.
To study the system-bath coupling in the Floquet picture, we can apply the same
unitary transition $U_F$, which block-diagonalizes $H_{\rm S,F}$, to the system
coupling operator. We then
obtain
\begin{equation}
 U_F^{\dagger} H_{\rm SB} U_F = \gamma
  \begin{pmatrix}
  \ddots & \vdots & \vdots & \vdots & \cdots \\
  \cdots & A_{\rm S, eff}^{(0)} & A_{\rm S, eff}^{(1)} & A_{\rm S, eff}^{(2)} & \cdots \\
  \cdots & A_{\rm S, eff}^{(-1)} & A_{\rm S, eff}^{(0)} & A_{\rm S, eff}^{(1)} & \cdots \\
  \cdots & A_{\rm S, eff}^{(-2)} & A_{\rm S, eff}^{(-1)} & A_{\rm S, eff}^{(0)} & \cdots \\
  \cdots & \vdots & \vdots & \vdots & \ddots \\
 \end{pmatrix}
 \otimes \sum_n c_n x_n,
\end{equation}
where $\hat{U}_F(t)^{\dagger}A_{\rm S}\hat{U}_F(t)=\sum_n A_{\rm S, eff}^{(n)}e^{-i n \omega t}$.

First of all, if only $A_{\rm S, eff}^{(0)} \neq 0$ and all other-order components
vanish, i.e. $A_{\rm S, eff}^{(n)} = 0$ for $n\neq 0$, the system-bath coupling operator
is also block-diagonal. This condition is equivalent to the case that 
$\hat{U}_F(t)^{\dagger}A_{\rm S}\hat{U}_F(t)=A_{\rm S, eff}^{(0)}$ does not depend on time.
In that case, we obtain a series of identical decoupled Hamiltonians
\begin{equation}
 H_{\rm eff} = H_{\rm S,eff}+ n\omega + H_{\rm B}+A_{\rm S, eff}^{(0)}\sum_n c_n x_n
\end{equation} 
($n= \cdots,-2,-1,0,1,2,\cdots$) in the block-diagonal parts.
Therefore, we can see that the statistical mechanics of Floquet states becomes the standard equilibrium
statistical mechanics of the effective Hamiltonian  $H_{\rm S,eff}$. 
(Note that equilibrium statistical mechanics may also be constructed using 
conserved Noether charge associated certain symmetry in full time-dependent Lagrangian \cite{Iadecola13}.)
The occupation of Floquet states follows the standard Boltzmann distribution 
(consider higher bath temperature), and one can define the concept of 
temperature for the system in a standard way. 
The order of the Floquet states and quasi-energies should be only considered
from the effective Hamiltonian eigenvalues.
If we have a weak system-bath coupling and Markovian conditions,  
the corresponding equilibrium statistical properties do not depend on the 
form of the coupling operator $A_{\rm S, eff}^{(0)}$. 

Secondly, if there exists $A_{\rm S, eff}^{(n)} \neq 0$ for $n \neq 0$ (i.e.
$\hat{U}_F(t)^{\dagger}A_{\rm S}\hat{U}_F(t)=A_{\rm S,eff}(t)$ depends on time),
one cannot simultaneously block-diagonalize both $H_{\rm S}$ and $A_{\rm S}$.
In that case, the off-diagonal blocks $A_{\rm S, eff}^{(n)}$ can induce
transitions between different Floquet equivalent sectors (e.g. between 
$H_{\rm S,eff}+n\omega$ and $H_{\rm S,eff}+m\omega$); and therefore, 
the Floquet picture is not exactly correct in this case. If we still consider
the statistical properties in the Floquet picture, the distribution becomes
complicated \cite{Kohler97,Breuer00,Kohn01,Hone09,Ketzmerick10}. 
It is unclear whether  one can find a universal theory in this regime.

{\em A Weak Version of the Classification Theorem}.
The classification scheme and their conditions in the last section are too abstract to be useful;
and therefore, we first look at an example and then extract some practical conditions.
We consider a non-interacting one-dimensional bosonic tight-binding chain
with a heat bath, and focus on the cases with periodic modulation of potential energy.
The system Hamiltonian is
\begin{eqnarray}
 H_{\rm S}(t)&=&\sum_{j=-M}^{M} \Big[ J (c_{j+1}^{\dagger}c_{j}+h.c.)
         -\mu n_{j} \nonumber \\
         && +V(j-\delta)^2 n_{j}\Big] +F(t)H_{\rm D},
         \label{eq:H_1Dchain}
\end{eqnarray}
where the operator $c_{j}^{\dagger}$ ($c_{j}$) creates (annihilates) a boson
on the $j-$site of the chain, $n_{j}=c_{j}^{\dagger}c_{j}$ is the number operator, and
the term $V(j-\delta)^2$ describes a quadratic potential on the chain
($\delta$ is an arbitrary constant). The last term is a time-periodic tilt 
potential where $F(t)=F(t+T)$ and $H_{\rm D}=\sum_{j} j n_{j}$ (define
$H_{\rm S}(t)=H_0+F(t)H_{\rm D}$). In continuum limit, the model above is equivalent to
a particle in one-dimensional potential well $H_{\rm S}(t)=P^2/(2m)+U(X,t)$, where
$P$ and $X$ are the momentum and position variables, $U(X,t)$ is time-periodic 
potential energy. For such a case, one usually
assume a linear system-bath coupling, i.e. $A_{\rm S}=X$; this results in a form
$A_{\rm S}=a\sum_{j} j n_{j}$ in the tight-binding model (where
$a=1$ is lattice constant). 

We can choose a time-dependent unitary transformation $\hat{U}_F(t)=e^{-if(t)H_{\rm D}}$,
and the function $f$ is chosen as
$df(t)/dt+F(t)=0$ and $f(t=0)=0$ \cite{eckardt05,Verdeny13}. By using this transformation,
one can approximately block-diagonalize the Floquet Hamiltonian of the Matrix form Eq.(\ref{eq:HSF_matrix}).
The off-diagonal blocks are not exactly zeros,
but in the large driving frequency limit $\omega\gg \{J,\mu,V(M+\delta)^2\}$
the transitions between different diagonal blocks can be treated perturbatively.
Up to the leading order of $\mathrm{max}[J,\mu,V(M+\delta)^2 ]/\omega$, one only keep 
the diagonal blocks, and obtain an effective Hamiltonian for the
system part \cite{eckardt05,Verdeny13}
\begin{equation}
 H_{\rm S, eff}=\!\!\sum_j \!\Big[ J (\mathcal{L}_{0} c_j^{\dagger}c_{j+1}+h.c.)
             -\mu n_{j} +V(j-\delta)^2 n_{j}\Big].
\end{equation}
where $\mathcal{L}_{0} = (1/T)\int_0^T dt e^{-if(t)}$.
Then, under a same transformation, we want to ask if the system-bath 
coupling operator can also be block-diagonalized, 
i.e. whether $\hat{U}_F(t)^{\dagger} A_{\rm S}\hat{U}_F(t)$ is independent of time.
This condition is equivalent to $[H_{\rm D},A_{\rm S}]=0$.
Now, the classification theorem for Floquet statistical mechanics can be 
weakened to:

``\textit{Assume a time-periodic system with a heat bath can be modeled by Eq.(\ref{eq:SB_Hamiltonian}).
In the large driving frequency limit, i.e. $\omega\gg D$ ($D$ is the largest energy scale, 
e.g. band width), if $[H_{\rm D},A_{\rm S}]=0$, the statistical mechanics
of the Floquet states is equivalent to the standard equilibrium statistical mechanics of 
the corresponding effective Hamiltonian $H_{\rm S,eff}$.}''

For example, if the system-bath coupling has a form $A_{\rm S}=\sum_{j} V_{SB}(j) \;(n_{j})^{\kappa}$ 
(where $V_{SB}(j)$ is an arbitrary function of site $j$, and $\kappa$ is an exponent.), 
any type of potential modulation 
(e.g. $H_{\rm D}=\sum_{j} j n_{j}$) satisfies the condition $[H_{\rm D},A_{\rm S}]=0$.
However, some types of modulations, e.g. periodic modulation of the hopping strength $J$, 
do not meet the condition. In addition, although we derive the theorem for one dimension boson chain,
the theorem can be applied to both boson and fermion in any dimension (note that the unitary transformation $U_F(t)$
in large frequency regime has the same form for all the cases).

\begin{figure}[t]
\centering
\vspace{0.05in} 
\includegraphics[width=3.3in,clip]{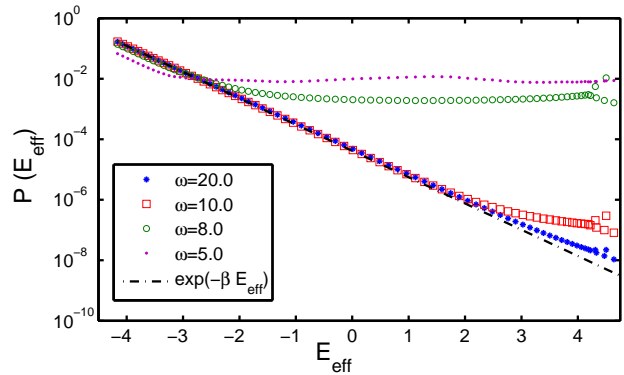}
\caption{(color online) The case $A_{\rm S}=\sum_{j} j n_{j}$.
The Floquet occupation distribution $P$ as a function
of the corresponding eigenvalues $E_{\rm eff}$ for the effective Hamiltonian $H_{\rm S,eff}$.
Different curves are for different driving frequency $\omega=20.0$ (star),
$\omega=10.0$ (square), $\omega=8.0$ (circle), and $\omega=5.0$ (dot).
The black dashed line is the exact Boltzmann distribution 
$P(E_{\rm eff})=e^{-\beta E_{\rm eff}}/\sum_{E_{\rm eff}} e^{-\beta E_{\rm eff}}$.
We choose $J=2.5$, $\mu=0.0$, $V=0.001$, $M=40$, $\delta=9.0$, $\beta=2.0$, 
$K\cdot T = 4.0$, $\omega=2\pi/T$, and cutoff $\epsilon_{c}=40\omega$.
} 
\label{fig:DensityBath}
\end{figure}

{\em Numerical Results from a Floquet-Markov Approach}.
Here, we will calculate the occupation distribution of the Floquet states for
the model in Eq. (\ref{eq:H_1Dchain}), and test the classification theorem.
With the heat bath, the occupation distribution is described by the reduced
density operator $\rho_{\rm S}(t)=Tr_{\rm B} \rho(t)$, where $\rho(t)$ is the density
operator for the whole system-bath model and $Tr_{\rm B}$ denotes a partial trace over the bath. 
For time-periodic system, one can 
adopt the Floquet-Markov approach \cite{Kohler97,Breuer00,Kohn01,Hone09,Ketzmerick10}:
1) The density matrix equation is simplified by the Markov approximation, which requires
a small bath correlation time compared to the relaxation time characterizing
the evolution of the system; 2) the master equation including the reduced density operator
is further projected onto the space of Floquet states:
\begin{eqnarray}
&&\rho_{\rm S(\alpha\beta)}(t)=\langle \phi_{\alpha}(t)|\rho_{\rm S}(t)|\phi_{\beta}(t)\rangle , \nonumber\\
&&A_{\rm S(\alpha\beta)}(t)=\langle \phi_{\alpha}(t)|A_{\rm S}|\phi_{\beta}(t)\rangle;
\end{eqnarray}
3) we consider the regime that the system-bath coupling strength is sufficiently small
compared to all the quasi-energy level spacings, so all the off-diagonal 
density matrix elements in the master equation can be neglected. With those approximations, the system has a
stationary solution for the occupation probability $P_{\alpha}=\rho_{\rm S(\alpha\alpha)}$ 
of Floquet state $|\phi_{\alpha}(t)\rangle$, which obey the rate equation \cite{Kohler97,Breuer00,Kohn01,Hone09,Ketzmerick10}
\begin{equation}
 0=P_{\alpha}\sum_{\beta} R_{\alpha\beta} - \sum_{\beta} P_{\beta} R_{\beta\alpha},
 \label{eq:masterEQ}
\end{equation}
and the normalization condition $\sum_{\beta} P_{\beta}=1$. 
Here, we show the main steps in solving the rate equation Eq. (\ref{eq:masterEQ}).
The rates describing the bath-induced transition 
between Floquet states are defined as
\begin{equation}
 R_{\alpha\beta}=2\pi \gamma^2 \sum_m |A_{\rm S(\alpha\beta)}(m)|^2 
            g(\epsilon_{\beta}-\epsilon_{\alpha}-m\omega),
\end{equation}
where $A_{\rm S(\alpha\beta)}(m)=\int_0^T dt e^{-i m \omega t} A_{\rm S(\alpha\beta)}(t)$.
The function $g(\epsilon)=n_{B}(\epsilon)J(\epsilon)/\pi$ is the correlation
function of the bath coupling operator $\sum_n c_n x_n$, where 
\begin{equation}
 n_{\beta}(\epsilon)=\left\{ 
  \begin{array}{l l}
    1/(e^{\beta\epsilon}-1) & \quad \text{if $\epsilon>0$}\\
    e^{-\beta\epsilon}/(e^{-\beta\epsilon}-1) & \quad \text{if $\epsilon<0$}
  \end{array} \right.
\end{equation}
is the Planck distribution for the bath with temperature $1/\beta$. 
The spectral density of the bath is
$J(\epsilon)=(\pi/2)\sum_n (c_n^2/m_n\omega_n)[\delta(\omega-\omega_n) - \delta(\omega+\omega_n))]$.
In the continuum limit for an ohmic bath with exponential cutoff, spectral density becomes
$J(\epsilon)\propto \epsilon e^{-|\epsilon|/\epsilon_{c}}$.

We numerically study the Floquet occupation distribution of a 1D finite chain ($M=40$)
modeled by Eq. (\ref{eq:H_1Dchain}) with square-wave modulation
\begin{equation}
F(t) = \left\{ 
  \begin{array}{l l}
    K & \quad \text{if $nT\leq t<(n+1/2)T$},\\
    -K & \quad \text{if $(n+1/2)T\leq t<(n+1)T$}.
  \end{array} \right.
\end{equation}
The quasi-energies and Floquet states can be obtained by solving
$U(T,0)|\phi_{\alpha}(0)\rangle=e^{-i\epsilon_{\alpha}T}|\phi_{\alpha}(0)\rangle$ and
$|\phi_{\alpha}(t)\rangle=e^{i \epsilon_{\alpha} t} U(t,0)|\phi_{\alpha}(0)\rangle$,
where $U(t,0)$ is the time evolution operator for the system.

\begin{figure}[t]
\centering
\includegraphics[width=3.3in,clip]{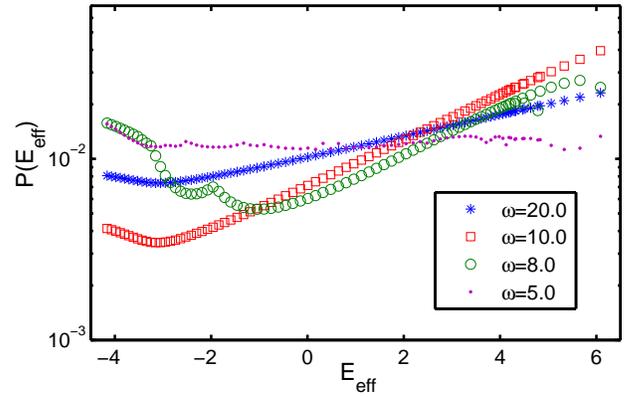}
\caption{(color online) The same as Fig.~\ref{fig:DensityBath}, but
for the case $A_{\rm S}=\sum_{j} c_{j+1}^{\dagger}c_{j} +h.c. $.
} 
\label{fig:HoppingBath}
\end{figure}

First of all, we consider a case  that the heat bath couples the particle density in the chain 
with the form $A_{\rm S}=\sum_{j} j n_{j}$ such that $[H_{\rm D},A_{\rm S}]=0$. 
In evaluating the transition rate $R_{\alpha\beta}$, it is necessary to truncate the summations:
We consider the summations from $m=-250$ to $250$ in numerics. 
By solving the rate equations in Eq. (\ref{eq:masterEQ})
with the normalization condition, one can obtain the Floquet occupation distribution $P$.
In Fig. \ref{fig:DensityBath}, we plot the Floquet occupation distribution as
a function of their corresponding eigenvalues $E_{\rm eff,\alpha}$ of $H_{\rm S, eff}$.
As mentioned before, the relation between $E_{\rm eff,\alpha}$ and quasi-energies is
$\epsilon_{\alpha}=\mod (E_{\rm eff,\alpha},\omega)$, where some non-zero off-diagonal terms in $U_F^{\dagger}H_{\rm S,F}U_F$ are neglected. 
For large driving frequency $\omega=20.0$ and $\omega=10.0$ (compared to bandwidth $D=4J=10.0$), the Floquet
occupation distribution is almost the same as Boltzmann distribution. As the driving
frequency decreases, the distribution starts to deviate from the Boltzmann distribution.
The reason is as follows. For smaller $\omega$, the energy separation between different
Floquet sectors is not large enough to prevent the bath-induced transitions among those 
sectors due to the presence of the off-diagonal blocks in 
$U_F^{\dagger}H_{\rm S,F}U_F$. Or we may also think the proposed unitary 
transformation $\hat{U}_F(t)$ is poor in those cases, 
and the correct transformation (which can block-diagonalize $H_{S,F}$)
cannot block-diagonalize the system-bath coupling operator, i.e.
$\hat{U}_F(t)^{\dagger}A_{\rm S}\hat{U}_F(t)$ is still time-dependent.

Secondly, we consider the system-bath coupling
$A_{\rm S}=\sum_{j} c_{j+1}^{\dagger}c_{j} +h.c. $, i.e. the bath couples to
the particle hopping processes
and so $[H_{\rm D},A_{\rm S}]\neq 0$. In that case, the Floquet occupation
distribution is shown in Fig. \ref{fig:HoppingBath}, and does not follow the
Boltzmann distribution even for large driving frequency. Interestingly, the distribution
first undergo an exponential decay and then becomes an exponential growth. 
It seems that one reaches an example with negative-temperature.
However, in the presence of the off-diagonal blocks $A_{\rm S, eff}^{(n)}$,
the equilibrium statistical mechanics for $H_{\rm S,eff}$ is meaningless
for the Floquet states. Those Floquet occupation distribution functions 
provide a different way to understand the order of the quasi-energies:
The larger Floquet occupation values correspond to the quasi-energies with ``lower positions''.
The general theory to correctly capture the statistical mechanics in this regime
is still waiting to be discovered.

{\em Discussions}.
We consider Floquet systems with the thermal bath coupling, and discover a classification theorem
for the statistical mechanics of the Floquet open systems. In large driving frequency, 
if the time-periodic part in the system Hamiltonian commutes with the system-bath coupling
operator, i.e. $[H_D,A_S]=0$, the statistical mechanics of the Floquet system can be described
by the standard equilibrium statistical mechanics of an effective Hamiltonian $H_{\rm S,eff}$.
The Floquet ``ground states'' correspond to the ground states of $H_{\rm S,eff}$, and the order of Floquet states
correspond to the order of eigenstates for $H_{\rm S,eff}$. However, if one has  $[H_D,A_S]\neq 0$,
the statistical distribution of the Floquet state becomes uncontrolled, and Floquet picture may not be the right way to
understand the statistical mechanics of time-periodic systems.
Therefore, our results have important implications for
engineering the Floquet systems and preparations of Floquet exotic states in realistic experimental systems.

D.E.L. is grateful to A. Levchenko and R. M. Lutchyn for valuable discussions.
D.E.L. acknowledges the support from Michigan State University in 
the problem formulation stage of the work.

\bibliography{FloquetETH_OPEN}

\end{document}